\documentclass[prd,reprint,amsmath,amssymb,preprintnumbers,superscriptaddress,nofootinbib]{revtex4-1}

\usepackage[utf8]{inputenc}
\usepackage{subfigure}
\usepackage{enumitem}
\usepackage{color}
\usepackage{graphicx}
\usepackage{amsmath}
\usepackage{hyperref}

\def\beq{\begin{equation}}
\def\eeq{\end{equation}}
\def\bea{\begin{eqnarray}} 
\def\eea{\end{eqnarray}}

\def\eps{\varepsilon}

\newcommand{\lsim}{
\mathrel{\hbox{\rlap{\hbox{\lower4pt\hbox{$\sim$}}}\hbox{$<$}}}}
\newcommand{\gsim}{
\mathrel{\hbox{\rlap{\hbox{\lower4pt\hbox{$\sim$}}}\hbox{$>$}}}}

\begin{document}
    
\preprint{CTPU-PTC-19-13}
\title{Implications of the dark axion portal for SHiP and FASER \\and the advantages of monophoton signals}

\author{Patrick deNiverville}
\affiliation{Center for Theoretical Physics of the Universe, IBS, Daejeon 34126, Korea}
\author{Hye-Sung Lee}
\affiliation{Department of Physics, KAIST, Daejeon 34141, Korea}
\date{April 2019}
\begin{abstract}
We investigate the implications of the dark axion portal interaction, the axion-photon-dark photon vertex, for the future experiments SHiP and FASER.
We also study the phenomenology of the combined vector portal (kinetic mixing of the photon and dark photon) and dark axion portal.
The muon $g-2$ discrepancy is unfortunately not solved even with the two portals, but the low-energy beam dump experiments with monophoton detection capability can open new opportunities in light dark sector searches using the combined portals.

\end{abstract}
\pacs{}
\maketitle

\section{Introduction}
\label{sec:Intro}
The existence of dark matter is one of the most important motivators for new physics scenarios.
The Standard Model (SM) can only explain about 5\% of the total energy budget of the Universe, while dark matter is expected to make up a much larger 27\% \cite{PhysRevD.98.030001}.
Although the weakly interacting massive particle (WIMP) picture has been a driving force in dark matter studies, the atmosphere has changed somewhat in the past decade due in part to the absence of signals after extensive searches by both underground and collider experiments \cite{Roszkowski:2017nbc}.
Among the various alternative pictures that have been put forward are that the dark sector could contain multiple different dark sector particles such as dark fermions, dark Higgs, the dark photon, and axions.

Communication between the dark sector and the SM can be established through the concept of a ``portal.''
For instance, the vector portal connects the SM photon to the dark photon through kinetic mixing \cite{Holdom:1985ag} and the axion portals such as the axion-photon-photon vertex connect the axion to two SM particles \cite{Sikivie:1983ip}.
These portals have been widely used by experiments to search for dark sector particles \cite{Essig:2013lka}.
Recently, a new portal called `dark axion portal' was suggested that connects both the axion and dark photon to the SM through axion-photon-dark photon as well as axion-dark photon-dark photon vertices, which can be exploited to study the axion and dark photon without relying on the two existing portals \cite{Kaneta:2016wvf}.

There have been several previous studies using this portal (for examples, see Refs.~\cite{Choi:2016kke,Kaneta:2017wfh,Agrawal:2017eqm,Alvarez:2017eoe,Daido:2018dmu,Choi:2018dqr,deNiverville:2018hrc,Huang:2018mkk}).
Of particular relevance to this work, Ref.~\cite{deNiverville:2018hrc} studied how measurements of lepton $g-2$ and experiments such as $B$-factories, fixed target neutrino experiments and beam dump experiments could be used to constrain the dark axion portal coupling $G_{a\gamma\gamma'}$.
The primary purposes of this paper are (i) to study the implications of the dark axion portal for the future experiments SHiP and FASER, and (ii) to explore new channels that become available when we consider both the vector portal and the dark axion portal.
For the latter, we discuss whether the $3.5 \sigma$ level muon $g-2$ anomaly \cite{PhysRevD.98.030001} can be explained in the presence of the two portals, and also point out substantial advantages in exploring the combined dark sector if monophoton searches can be implemented in beam dump experiments.

The rest of the paper is organized as follows.
In Sec.~\ref{sec:DAP}, we provide a short overview of the dark axion portal and define our parametrization.
In Sec.~\ref{sec:ship}, we calculate the projected sensitivity of the SHiP experiment using CERN's SPS proton beam line.
In Sec.~\ref{sec:faser}, we calculate the projected sensitivity of the FASER and FASER 2 experiments at the LHC.
In Sec.~\ref{sec:other_experiments}, we study three experiments, MATHUSLA, NA62 and REDTOP, which we do not project to provide new constraints on the dark axion portal with their currently planned analyses.
In Sec.~\ref{sec:muon_g_2}, we show that a combination of the vector and dark axion portals is incapable of explaining the muon $g-2$ discrepancy due to constraints on the coupling strengths of the two portals.
In Sec.~\ref{sec:dm}, we consider how the inclusion of an invisible decay channel to some invisible dark matter candidate $
chi$, $\gamma^\prime \to \chi \bar \chi$, affects muon $g-2$ and the limits discussed in this paper.
In Sec.~\ref{sec:monophoton}, we discuss how a beam dump experiment exploiting the monophoton signal may be used to probe a combination of the dark axion and vector portals. 
In Sec.~\ref{sec:summary}, we summarize the results.

\section{Dark axion portal}
\label{sec:DAP}
The axion and dark axion portals introduce the following new interaction terms,
\begin{eqnarray}
&& {\cal L}_\text{axion portal} = \frac{G_{agg}}{4} a G_{\mu\nu}\tilde G^{\mu\nu} + \frac{G_{a\gamma\gamma}}{4} a F_{\mu\nu}\tilde F^{\mu\nu} + \cdots ~~ \\
&&{\cal L}_\text{dark axion portal} = \frac{G_{a\gamma\gamma^\prime}}{2} a F_{\mu\nu}\tilde Z'^{\mu\nu} + \frac{G_{a\gamma^\prime\gamma^\prime}}{4}  a Z'_{\mu\nu}\tilde Z'^{\mu\nu}  ~~~ 
\end{eqnarray}
where $a$ stands for the axion (or axion-like particle), and $G_{\mu\nu}$, $F_{\mu\nu}$, $Z'_{\mu\nu}$ stand for the gluon ($g$), photon ($\gamma$), and dark photon ($\gamma'$) field strengths, respectively.
The axion and dark axion portals are produced through anomaly triangles although the exact couplings are dependent on the underlying model. An example implementation through the dark KSVZ model can be found in Ref. \cite{Kaneta:2016wvf}.

The vector portal is compatible with the dark axion portal, and introduces a kinetic mixing (parametrized by $\eps$) between the photon and the dark photon,
\begin{equation}
 {\cal L}_\text{vector portal} = \frac{\eps}{2} F_{\mu \nu} Z^{\prime \mu\nu}.
\end{equation}

We will assume the same setup as Ref.~\cite{deNiverville:2018hrc}, that is that $m_a \ll m_{\gamma^\prime}$, and that the model-dependent parameters are arranged such that $G_{a\gamma\gamma}$ is sufficiently small that its effects can be neglected in the processes considered in this paper. The small mass $m_a$ ensures that the $a$ is long lived on the timescales considered by beam dump and fixed target experiments. We will focus on the effects of the new $G_{a\gamma\gamma^\prime}$ coupling introduced by the dark axion portal in the regime where kinetic mixing $\eps=0$ in Secs. \ref{sec:ship} and \ref{sec:faser} and consider the case where both portals are open afterward.

It is worthwhile to mention that the $G_{a\gamma\gamma'}$ coupling could potentially be constrained directly or indirectly by channels we do not consider in this paper;
directly by considering the astroparticle phenomena such as the stellar cooling or horizontal branch stars; indirectly by the relation between the $G_{a\gamma\gamma}$ and $G_{a\gamma\gamma'}$ depending on the model.
In this paper, however, we limit ourselves only to controlled laboratory experiments and study what they tell us about the direct constraints and implications of the new portal.

\begin{figure}[t]
 \centerline{\includegraphics[width=0.48\textwidth]{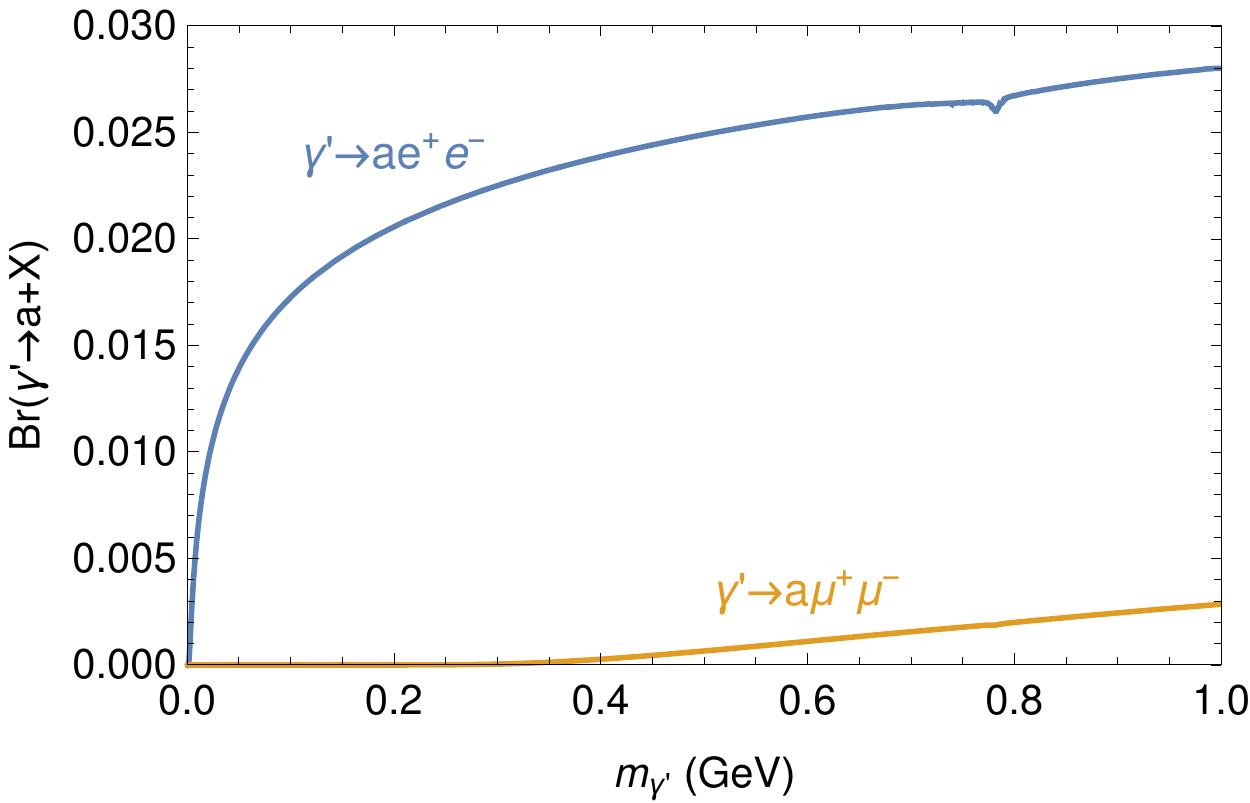}}
 \caption{The branching ratio of $\gamma^\prime \to a e^+ e^-$ and $\gamma^\prime \to a \mu^+ \mu^-$ through the dark axion portal in the $m_a \ll m_{\gamma'}$ limit. The leptonic decay processes are forbidden for $m_{\gamma^\prime} < 2 m_\ell$ $(\ell=e, \mu)$. The slight dip near 0.8\,GeV is due to the presence of hadronic decay channels.}
 \label{fig:dp_to_aee}
\end{figure}

The dark photon $\gamma^\prime$ may decay through a number of different channels in the dark axion model. The two-body decay $\gamma^\prime \to a\gamma$ through the dark axion portal is the dominant channel with a decay width
\begin{equation}
 \Gamma(\gamma^\prime\to\gamma a) = \frac{G_{a\gamma \gamma^\prime}^2}{96\pi} m_{\gamma^\prime}^3\left(1-\frac{m_a^2}{m_{\gamma^\prime}^2} \right)^3.
\end{equation}
The three-body decay processes $\gamma^\prime \to e^+ e^- a$ and $\gamma^\prime \to \mu^+ \mu^- a$ are also possible, with a branching fraction of a few percent that increases with $m_{\gamma^\prime}$, as shown in Fig. \ref{fig:dp_to_aee}. While not dominant, these channels provide a signature very similar to the lepton antilepton signal used to search for kinetically mixed visibly decaying dark photons, and those searches can be repurposed as probes of the dark axion portal.

\section{SHiP}
\label{sec:ship}

The search for hidden particles (SHiP) \cite{Alekhin:2015byh,Anelli:2015pba} is a proposed proton beam dump experiment using the CERN SPS. They plan to impact $2\times 10^{20}$ protons with 400 GeV of energy onto a molybdenum target over five years of running. The current plans call for both a neutrino detector capable of searching for $\nu_\tau$ interactions and a decay pipe followed by electron and hadron calorimeters to search for the decays of rare, long-lived particles. This analysis will focus on dark axion portal signals visible in the SHiP decay volume: electron-positron pair production and a monophoton signal. 

We will follow the $\gamma^\prime$ search of Ref. \cite{Gorbunov:2014wqa} with some input from Refs. \cite{Graverini:2214085,SHIPDP} for a possible decay volume and detector geometry. We assume a 50-meter long cylindrical decay pipe with a diameter of 5 meters. The upstream face of the cylinder is located 50 meters from the production target and is parallel to the beam line with no horizontal or vertical offset.

\begin{figure}[t]
\includegraphics[width=0.48\textwidth]{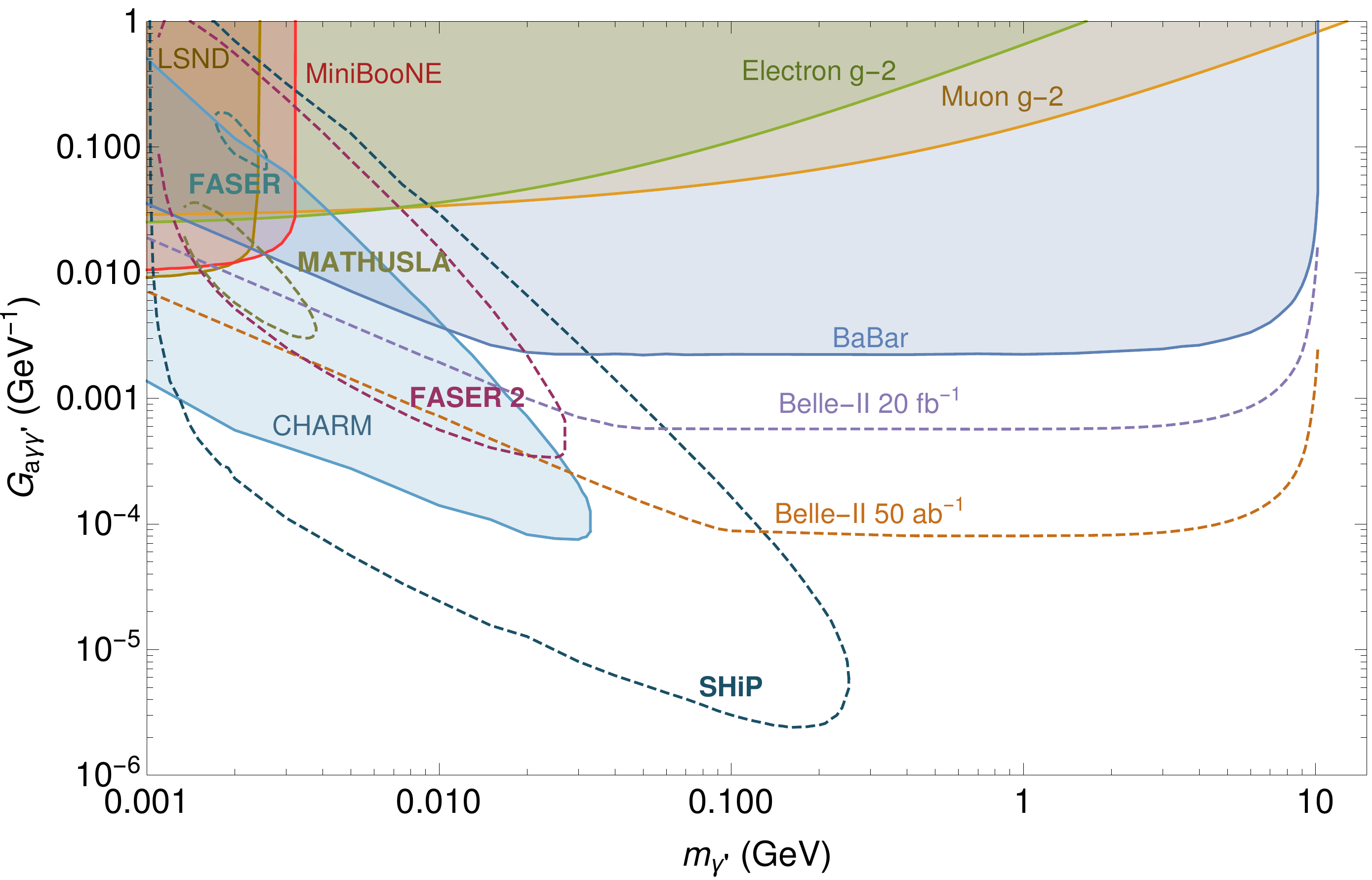}
\caption{Limits on $G_{a\gamma\gamma^\prime}$ from high intensity experiments and the projected sensitivity of the SHiP, FASER and MATHUSLA experiments for $m_a \ll m_{\gamma^\prime}$ with $\varepsilon = 0$. The limits from LSND and MiniBooNE come from excess neutral current-like elastic scattering events above the expected background from neutrino and non-neutrino SM sources. The LSND limit reflects an excess of 2$\sigma$ over expected backgrounds, while MiniBooNE assumes zero background and excludes the scenario at a 90\% confidence level. The CHARM constraint reflects sensitivity to a 2$\sigma$ excess in monophoton production through $\gamma^\prime \to a\gamma$ decays in the CHARM fine-grain detector. The electron and muon $g-2$ lines indicate where the scenario would degrade the agreement between theory and experiment by more than $2\sigma$. The \textsc{BABAR} and Belle-II lines all represent exclusions from 2$\sigma$ monophoton excesses through the annihilation process $e^+ e^- \to a (\gamma^\prime \to a \gamma)$. Further details on these limits can be found in Ref. \cite{deNiverville:2018hrc}. The SHiP projection denotes a 95\% confidence level excess of greater than three $e^+ e^-$ events reaching the SHiP electromagnetic calorimeter produced through the decay $\gamma^\prime \to a e^+ e^-$, while the FASER and FASER 2 projections reflect the same for three events. Finally, the green MATHUSLA limit denotes an excess of greater than four observed events.}
\label{fig:ship_limits}
\end{figure}

Following previous work in modeling dark axion portal production at the CHARM experiment in Ref. \cite{deNiverville:2018hrc}, we consider production through the decays of the pseudoscalar mesons $\pi^0$ and $\eta$. The neutral pseudoscalar mesons are produced in large quantities at fixed target and beam dump experiments, though rarely studied as they do not supply significant background or signal without the introduction of new physics. The rare decays through the dark axion portal  $\pi^0,\eta\to a \gamma \gamma^\prime$ provide an important source of dark photons whose visible decay products could be detected in beam dump experiments like SHiP. We could also consider production through bremsstrahlung, though the required $2\to 4$ production process is expected to suppress the overall production rate significantly.

The simulation of the production, propagation, and decay of dark axion portal particles was performed with a modified version of the \textsc{BdNMC} software package \cite{deNiverville:2016rqh}. The generation of pseudoscalar meson decays requires knowledge of the angular-momentum production distributions for both the $\pi^0$ and $\eta$, as well as an estimate of the overall production rate. A number of different methods of simulating these meson distributions were recently tested and compared with available experimental data in the very helpful review included in Ref. \cite{Dobrich:2019dxc} in the context of axion-like particle production. One approach that has been employed previously is to approximate the $\pi^0$ distribution by the mean of a pair of appropriate $\pi^+$ and $\pi^-$ distributions, as this closely matches the expected $\pi^0$ momentum distribution at energies far above the pion mass \cite{Amaldi:1979zk,Jaeger:1974pk}. The $\eta$ distribution is quite similar to the $\pi^0$ distribution when it is above threshold production energies, and we will use the same distribution for both particles. For SHiP energies, we adopt the charged pion distributions of Ref. \cite{Bonesini:2001iz}, collectively denoted the BMPT distribution, as they are suitable for a variety of target materials and beam energies.

The overall $\pi^0$ production rate $N_{\pi^0}$ was estimated to be approximately 2.7 per proton on target (POT) in Ref. \cite{SHIPDP}. Note that the number of $\gamma^\prime$ per POT normalized by $\varepsilon^{-2}$ was calculated to be 5.41. The branching ratio of $\pi^0\to \gamma \gamma^\prime$ is $2\varepsilon^2 \times\left(1-\frac{m_{\gamma^\prime}^2}{m_{\pi^0}^2}\right)^3$ \cite{Batell:2009di}, and neglecting the mass dependent phase space factors, we can divide by two to obtain the estimated number of $\pi^0$s per POT. This estimate is slightly larger than but not inconsistent with those used previously in Refs. \cite{Alekhin:2015byh,deNiverville:2016rqh}, and is quite conservative when compared to several of the production estimates of Ref. \cite{Dobrich:2019dxc}. The $\eta$ production rate is scaled to that of the $\pi^0$. For the SHiP energy, we take $N_\eta = 0.1 \times N_{\pi^0}$ \cite{SHIPDP}. 

Sample meson 4-momenta are generated from the BMPT distribution using a simple acceptance-rejection algorithm. As the lifetime of the $\pi^0$ and $\eta$ are extremely short at $\mathcal{O}(10^{-17}\,\mathrm{s})$ or less \cite{PhysRevD.98.030001}, they do not propagate a significant distance before decaying. We simulate the three-body decay $\pi^0,\eta\to a\gamma\gamma^\prime$ as described in Ref. \cite{deNiverville:2018hrc}, discarding the $\gamma$ and $a$ 4-momenta as they do not contribute to the SHiP decay signal. The resulting list of $\gamma^\prime$ 4-momenta is used to calculate the expected dark axion portal signal. 

The probability that a $\gamma^\prime$ with label $i$ decays inside the SHiP decay pipe through an observable channel is given by
\begin{equation}
\label{eq:dec_prob}
 P_\mathrm{decay,i} = \mathrm{Br}_X \left[\exp\left(-\frac{L_{1,i} E_i}{c\tau m_{\gamma^\prime}}\right) - \exp\left(-\frac{L_{2,i} E_i}{c\tau m_{\gamma^\prime}}\right)\right],
\end{equation}
where\begin{itemize}
      \item $X=\gamma^\prime\to a e^+ e^-$,
      \item $E_i$ is the energy of the $\gamma^\prime$,
      \item $\tau$ is the lifetime of a $\gamma^\prime$ with mass $m_{\gamma^\prime}$ and coupling strength $G_{a\gamma\gamma^\prime}$,
      \item $L_{1,i}$ is the distance the $\gamma^\prime$ propagates before entering the decay volume, and
      \item $L_{2,i}$ is the total distance traveled before exiting the decay volume.
     \end{itemize}
     
Each $\gamma^\prime$ is decayed into an $a e^+ e^-$ final state. In order for this state to be accepted, both leptons must intersect with the end cap of the decay volume in order to enter the SHiP calorimeters, and the momenta must satisfy $p_{e^+},p_{e^-}>1\,\mathrm{GeV}/c$ \cite{Gorbunov:2014wqa}. The total event rate expected from the decays of meson $j$ can be calculated as
\begin{equation}
 \label{eq:decay_event_rate}
 N_{\mathrm{event},j} = \frac{N_j \epsilon_\mathrm{eff}}{N_\mathrm{trials}} \mathrm{Br}(j\to a \gamma \gamma^\prime) \sum_i P_\mathrm{decay,i} \theta(p_{e^+,i},p_{e^-,i}),
\end{equation}
where \begin{itemize}
 \item $j=\pi^0,\eta$,
 \item $P_\mathrm{decay,i}=0$ if the $\gamma^\prime$ does not intersect the detector,
 \item $\theta(p_{e^+,i},p_{e^-,i})=1$ if the end-state leptons satisfy the cuts mentioned above and 0 otherwise,
 \item $\epsilon_\mathrm{eff}=1$ is the detection efficiency, and
 \item $N_\mathrm{trials}$ is the total number of $\gamma^\prime$ trajectories generated.
\end{itemize} The total event rate is found by summing over $j$.

The projected sensitivity curve in Fig. \ref{fig:ship_limits} was generated by scanning the dark axion parameter space in $G_{a\gamma\gamma^\prime}$ and $m_{\gamma^\prime}$. We assume zero background, and therefore exclude the scenario when the predicted number of events generated $N_\mathrm{event}>3$. Note that the signal weakens considerably as $m_{\gamma^\prime}\to 2 m_e$, as the decay width of $\gamma^\prime \to a e^+ e^-$ becomes heavily suppressed by the available phase space, as shown in Fig. \ref{fig:dp_to_aee}. The branching ratio suppression is not reflected in the otherwise similar CHARM limit, as it instead considered a monophoton signal produced through $\gamma^\prime \to a\gamma$. Despite sizable backgrounds, the monophoton search at CHARM was considerably more sensitive than a search for $e^+ e^-$ final states. 

\begin{figure}[t]
\includegraphics[width=0.48\textwidth]{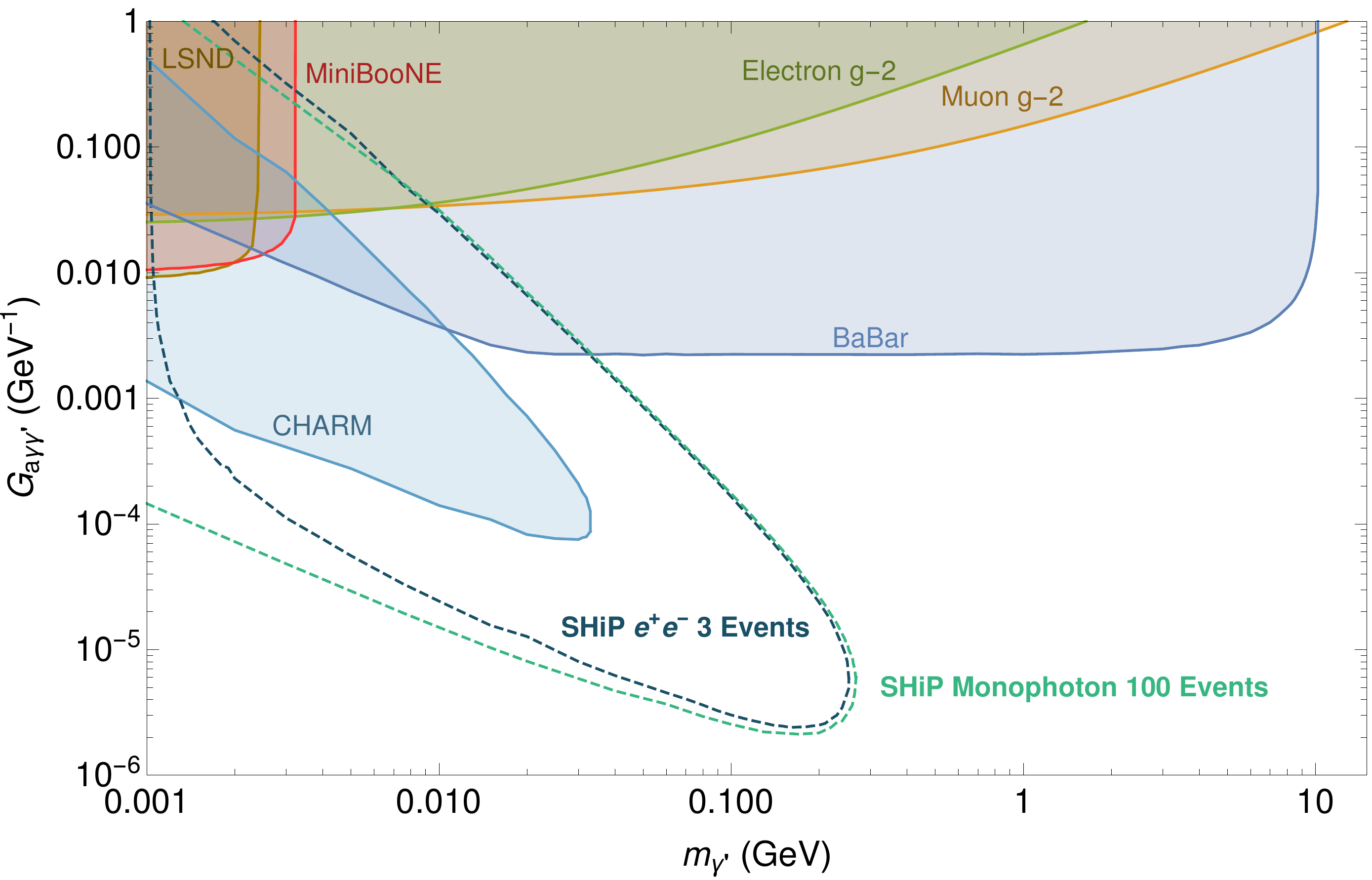}
\caption{The expected shape of a SHiP monophoton exclusion contour, with a required event rate of 100 chosen for comparison with the electron-positron signal. The monophoton search does not suffer any suppression in its sensitivity for $m_{\gamma^\prime} \approx 2 m_e$, and is therefore capable of greatly improving upon the electron-positron signal at low masses. The 100-event contour matches or surpasses the zero-background electron-positron contour over the entire parameter space of interest. Note that this is not a projection of sensitivity, as the actual backgrounds could be substantially worse and will require study by the collaboration to estimate.}
\label{fig:ship_mono}
\end{figure}

SHiP may benefit from a similar monophoton search, but this would require an estimate of the expected monophoton backgrounds, the analysis of which is currently in progress. Such a search would need to reject radiative processes from long-lived muons produced at the interaction point or through downstream neutrino interactions with matter. The rare decay $\mu \to e \nu_\mu \nu_e \gamma$ could be particularly problematic if the detector does not properly identify the electron. If the temporal and spatial resolution of reconstructed monophotons is poorer than that of charged particles, then coincident cosmic rays and their products can also pose a problem. To illustrate the potential of a SHiP monophoton search, we have simulated the potential signal, requiring that the emitted photon intersects with the end of the SHiP decay pipe and possess energy larger than 2\,GeV to match the combined energy of the electron and positron. In Fig. \ref{fig:ship_mono}, we show the contour where SHiP would observe 100 events, as was used in the CHARM monophoton analysis. Note that the SHiP monophoton contour does not suffer from the suppression in signal strength at $m_{\gamma^\prime} \approx 2 m_e$ exhibited by the SHiP electron-positron signal. Should studies indicate that the SHiP experiment is capable of a background free search for monophotons, though, the monophoton search will provide a dramatic improvement in sensitivity over the $e^+ e^-$ channel.

There are a few interesting comments to be made about this search:
\begin{itemize}
 \item We have assumed that the $a$ possesses a lifetime sufficiently long to not decay on the timescale required to travel from the SHiP target to the SHiP decay volume, but this is not essential to the analysis. There is no requirement that the $a$ produced through meson decay survive to reach the detector, and the secondary $a$ produced through the $\gamma^\prime \to a e^+ e^-$ would result in a diphoton coincident with the $e^+ e^-$ signal. The additional particles could either be ignored or an additional analysis could search for this more complicated end state. A search for $\gamma^\prime \to a \gamma$ would also need to consider a three-photon final state produced through the axion portal decay $a\to \gamma\gamma$. The advantage of these more complex final states is that all of the 4-momentum would be accounted for, rendering it possible to reconstruct the mass of the $\gamma^\prime$ if the uncertainties in the measured energies are sufficiently small while simultaneously providing a window into the interactions of the $a$.
 \item We did not consider the $a \mu^+ \mu^-$ final state, nor the possibility of bremsstrahlung $\gamma^\prime$ production through the dark axion portal, but their inclusion may extend SHiP's sensitivity to slightly larger values of $m_{\gamma^\prime}$. 
\end{itemize}

\section{FASER}
\label{sec:faser}

The forward search experiment (FASER) \cite{Ariga:2018zuc,Ariga:2018uku,Ariga:2018pin} is a proposed experiment with sensitivity to weakly coupled, long-lived particles at the LHC. Searches for long-lived particles with high transverse momenta are difficult at the LHC due to the small Standard Model production cross sections for such particles. By placing a detector in the far-forward region of an existing LHC interaction point, FASER can search for low transverse momentum, long-lived particles for which the production cross section is much larger. The planned location of FASER is 480\,m downstream from an LHC interaction point, by which point the beam has curved away, and the intervening rock and dirt have removed most SM particles. FASER would then search for the visible decays of weakly coupled long-lived particles, taking advantage of the extremely high boost from the high energy of the LHC to extend their lifespans.

The FASER collaboration has performed a preliminary analysis of its sensitivity to visible dark photon decays in Refs. \cite{Feng:2017uoz,Ariga:2018uku}, and we will be adopting these searches to the dark axion portal. The inelastic proton-proton cross section was measured to be $\sigma_\mathrm{inelastic} \sim 75\,\mathrm{mb}$ during the 13 TeV LHC run, and is not expected to differ greatly for the 14 TeV collisions of LHC run 3. The total number of inelastic collisions is therefore expected to be $N_\mathrm{inelastic} \approx 1.1\times 10^{16}$ for 150\,fb$^{-1}$ in LHC run 3.

As with the beam dump experiments studied previously, we expect $\pi^0$ and $\eta$ decays to provide the primary source of dark axion portal particles. The $\pi^0$ and $\eta$ production rates and distributions were generated with EPOS-LHC \cite{Pierog:2013ria} through the CRMC v1.7 framework \cite{crmc:2019}. Note that these rates were also compared with SIBYLL v2.3 and found to be consistent at the small angles required by FASER \cite{Ahn:2009wx,Riehn:2015oba}. The total number of $\pi^0$s ($\eta$s) per interaction in one hemisphere of the interaction point was calculated to be 19 (2.1). The production estimate is conservative, as we would also expect secondary meson production to result from other collision products impacting on material between the interaction point and the FASER detector.

\begin{figure}[t]
   \centerline{\includegraphics[width=0.48\textwidth]{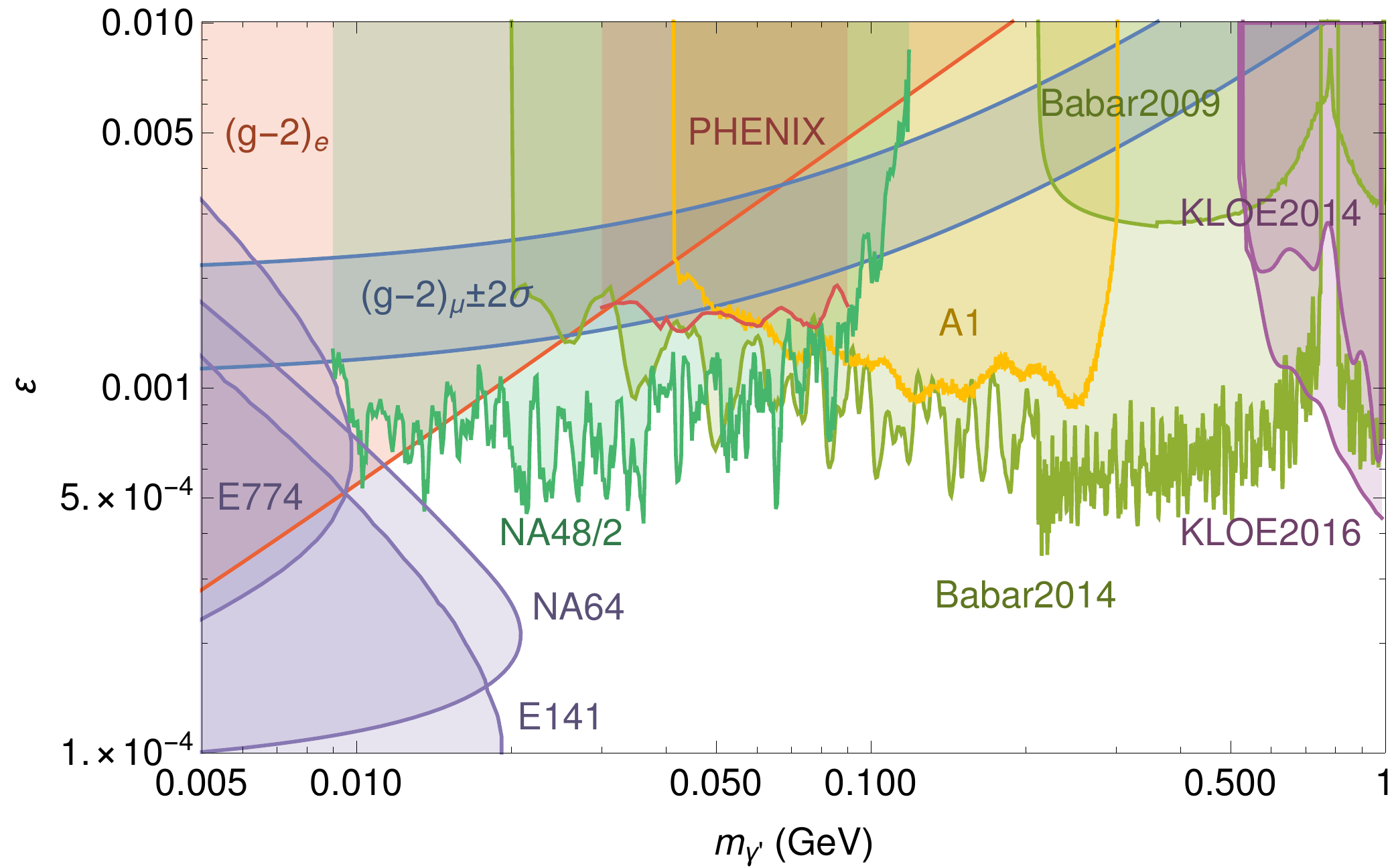}}
 \caption{Leading constraints on visibly decaying dark photon $\gamma^\prime$ coupling to the Standard Model through the vector portal. Limits are at a 90\% confidence level and given in terms of the kinetic mixing parameter $\varepsilon$. The blue shaded region shows where a dark photon can correct $(g-2)_\mu$ to within 2$\sigma$ of its measured value \cite{Mohr:2012tt,Bennett:2002jb,Bennett:2004pv,Bennett:2006fi,PhysRevD.98.030001}. Shown are the limits from \textsc{BABAR} \cite{Aubert:2009cp,Lees:2014xha}, KLOE-2 \cite{Babusci:2014sta,Anastasi:2016ktq}, PHENIX \cite{Adare:2014mgk}, A1 \cite{Merkel:2014avp}, NA48/2 \cite{Batley:2015lha}, E774 \cite{Bross:1989mp} and E141 \cite{Riordan:1987aw} from Ref. \cite{Bjorken:2009mm}, NA64 \cite{Banerjee:2018vgk} and electron $g-2$ \cite{Endo:2012hp}. Relevant limits suppressed for readability include two additional limits from KLOE-2 \cite{Archilli:2011zc,Babusci:2012cr}, WASA-at-COSY \cite{Adlarson:2013eza} and HADES \cite{Agakishiev:2013fwl}. Appearing at larger masses than are visible on this plot are those produced by LHCb \cite{Aaij:2017rft}. Several of the curves used in the making of this plot were taken from the darkcast data files \cite{Ilten:2018crw, darkcast}.}
 \label{fig:vector_limits}
\end{figure}

For this analysis, FASER is assumed to be a 1.5\,m long cylindrical decay region with 10\,cm radius located 480\,m from the interaction point operating during LHC run 3. We will also consider the sensitivity of a hypothetical FASER 2 detector, a 5\,m cylinder with 1\,m radius located 480\,m downstream from the interaction point. FASER 2 would take data during the high luminosity LHC runs with an expected luminosity 20 times larger than that of LHC run 3.

FASER is sensitive to the dark axion portal decay $\gamma^\prime \to a e^+ e^-$, as it possesses a signature very similar to that of the kinetically mixed dark photon. We will follow the cuts imposed by the FASER dark photon analysis: The dark photon must decay in the decay volume, both the electron and positron must cross through the downstream face of the decay volume, and the total visible energy must satisfy $E_{e^+}+E_{e^-}>100\,\mathrm{GeV}$. If we only consider mesons with energies greater than $100\,\mathrm{GeV}$, the number of $\pi^0$s ($\eta$s) per POT drops to 2.43 (0.43) in the hemisphere facing FASER. 

We assume negligible background and therefore impose a cut on the parameter space predicted to generate more than three events. We show the resulting contours in Fig. \ref{fig:ship_limits}, where the small inner contour represents the sensitivity of FASER and the larger outer contour that of FASER 2. The comparatively low luminosity hampers the ability of FASER to probe the scenario compared to beam dump experiments, and only FASER 2 is capable of excluding new parameter space. Improving on this search with a monophoton analysis would be challenging, as photon backgrounds are expected to be significant, and CHARM already excludes the region where the most improvement is expected (as shown in a hypothetical SHiP monophoton search in Fig. \ref{fig:ship_mono}). 

\section{Other Experiments}
\label{sec:other_experiments}

In this section, we discuss three proposed experiments (MATHUSLA, REDTOP, and NA62) mentioned in Ref. \cite{Beacham:2019nyx} that we do not expect to place significant novel limits on the dark axion scenario with their current analyses. Adjustments to each are possible to improve their sensitivity to the dark axion portal.

\subsection{MATHUSLA}
\label{ssec:mathusla}

MATHUSLA (massive timing hodoscope for ultra-stable neutraL particles) \cite{Curtin:2017izq,Curtin:2018mvb} is a proposed detector with many similarities to FASER in its objective, but with a very different approach to detector design and position. The proposed detector is a massive decay chamber $200\times200\times20\,\mathrm{m}^3$ in volume, with five resistive plate capacitor tracking layers spaced over an additional five meters above the decay volume. The detector complex is located 100\,m above and 100\,m downstream of the interaction point.\footnote{This places the center of the decay volume 110\,m above and 200\,m downstream of the interaction point.}

Our calculation of MATHUSLA's sensitivity is very similar to that of FASER 2: We use EPOS-LHC to estimate the production distribution and rate of $\pi^0$ and $\eta$ mesons. As MATHUSLA expects to be sensitive to electrons with energies of greater than 1 GeV, we impose a cut of $E_{\pi^0,\eta}<2\,\mathrm{GeV}$, and find estimates of
\[
 N_{\pi^0} \approx 5.5\times10^{18}\quad\mathrm{and}\quad N_\eta \approx 0.74\times 10^{18}
\]
for the HL-LHC run. Note that this is smaller by a factor of 3 than MATHUSLA estimates, but using the larger number would not change our conclusions. Also, we require that both the electron and positron cross all five tracking layers with a minimum separation of 1\,cm. We follow the MATHUSLA estimates and place an exclusion on four events while assuming approximately $100\%$ efficiency.

We only consider the $\gamma^\prime \to a e^+ e^-$ final state as it is not yet known whether the MATHUSLA detector will be built in such a way as to be sensitive to photons, and any monophoton modes would possess more challenging backgrounds \cite{Curtin:2018mvb}. Despite the lenient cuts and low expected backgrounds, MATHUSLA does not appear to possess new sensitivity to the dark axion portal, only probing parameter space already covered by CHARM. Approximately 10\% of events survive the energy cuts due to the relatively low energy of the off-axis $\pi^0$ and $\eta$ particles that intersect the MATHUSLA detector. We note that MATHUSLA does improve significantly with more generous assumptions on the overall production rate such as the larger $N_\pi^0$ estimate reported in the Ref. \cite{Curtin:2018mvb}, but we do not expect it to escape the parameter regime excluded by CHARM.

\subsection{REDTOP}
\label{ssec:redtop}

REDTOP\footnote{\url{https://redtop.fnal.gov/wp-content/uploads/2016/02/REDTOP_EOI_v10.pdf}} (rare eta decays with a TPC for optical photons) is a proposed $\eta$ factory with plans to deliver $10^{18}$ 1.8\,GeV POT and a detector with very high solid angle coverage. This experiment can place impressive new limits on visible dark photon decays through the process $\eta \to \gamma A^\prime \to \gamma e^+ e^-$. While this decay has a reasonably large SM branching ratio ($\sim7\times10^{-3}$), a bump search of the $e^+ e^-$ invariant mass can reveal the presence of the dark photon. This is less effective in the dark axion scenario, as not only is the branching ratio of the equivalent process $\eta \to a \gamma^\prime \gamma$ heavily suppressed relative to the minimal dark photon but the $\gamma^\prime$ must decay to the semivisible three-body final state $a e^+ e^-$, dramatically complicating any attempt to reconstruct its mass. Should REDTOP place strong limits on $\eta \to \gamma e^+ e^- + \mathrm{missing\,energy}$, it may be able to constrain the dark axion portal in the region between the current \textsc{BABAR} and CHARM limits.

\subsection{NA62}
\label{ssec:na62}

NA62 \cite{NA62:2017rwk,Dobrich:2018ezn,Dobrich:2019dxc} is a charged kaon decay experiment using the 400 GeV proton beam at the CERN SPS. We take the decay volume to be a 135 meter long cylinder beginning 82 meters downstream of the target with a 1 meter radius. Only the first 75 meters are considered for charged decays, as we force the leptons to cross the first spectrometer chamber in order to guarantee that particles are correctly identified. Both leptons must cross the liquid krypton calorimeter (LKr) at the end of the decay pipe with a spatial separation of at least 10 cm. Both leptons must also be at least 15 from the central hole of the LKr. We simulated the NA62 experiment for $10^{18}$ POT and assumed 100\% efficiency. Unfortunately, the dark axion portal rarely satisfies the separation condition as $\gamma^\prime \to a e^+ e^-$ frequently results in a highly collimated lepton pair. Were this cut less severe, NA62 would be likely to impose some limits on the dark axion parameter space.

\section{Muon $G-2$ Reconsidered}
\label{sec:muon_g_2}

\begin{figure}[t]
   \centerline{\includegraphics[width=0.48\textwidth]{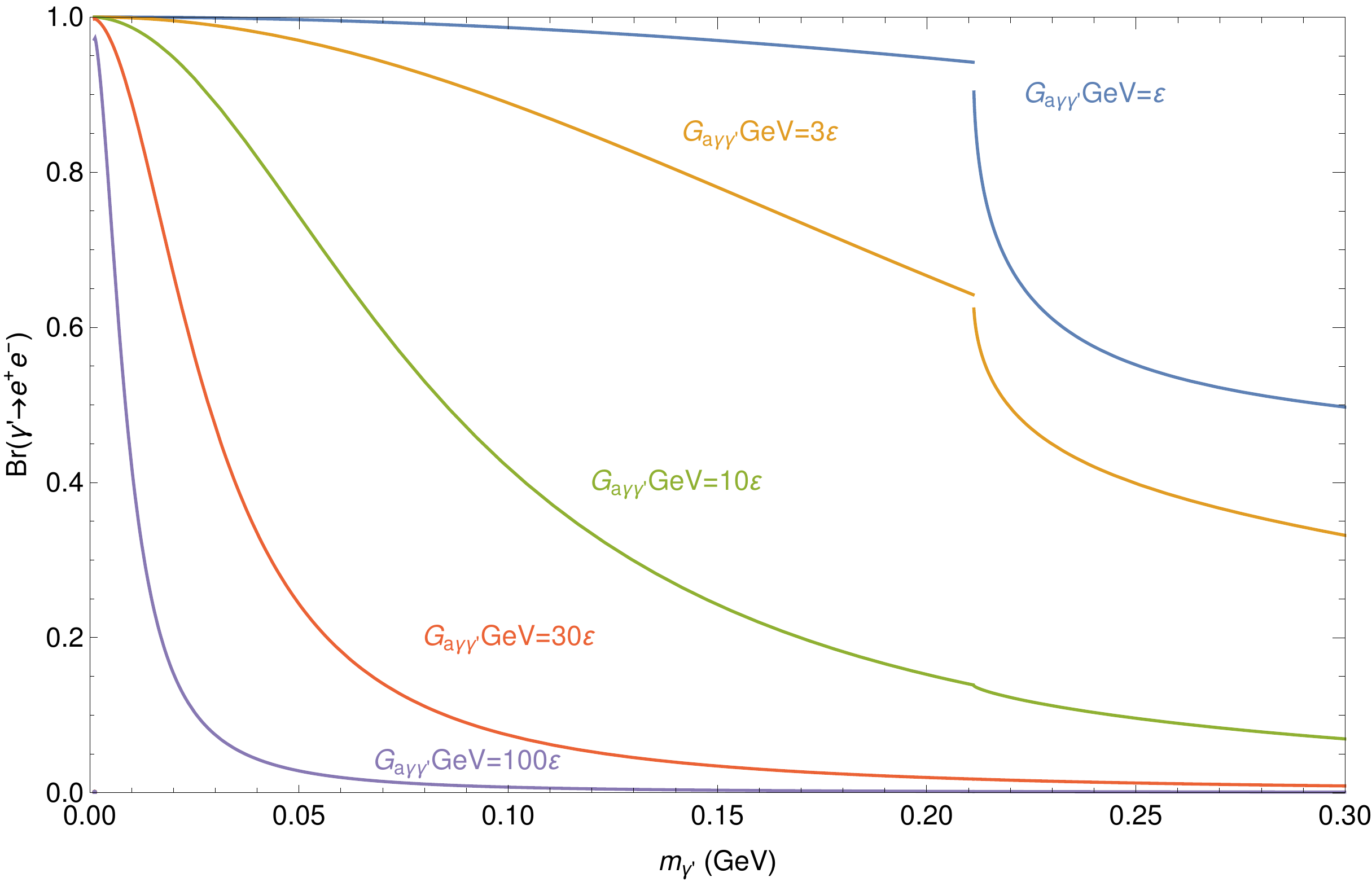}}
 \caption{The branching ratio of the vector portal process $\gamma \to e^+ e^-$ for five different ratios of $G_{a\gamma\gamma^\prime}$ to $\varepsilon$. The kink at approximately 210 MeV is due to the $\gamma^\prime\to\mu^+ \mu^-$ channel. For small $m_{\gamma^\prime}\ge 2 m_e$, the kinetic mixing decay dominates due to the weaker dependence of \ref{eq:dec_kinetic} on the mass of the dark photon. At $m_{\gamma^\prime}\le 2 m_e$, $\gamma^\prime\to e^+ e^-$ is forbidden by energy conservation, and this appears in the plot as a nearly vertical line next to the y axis where the branching ratio drops to zero.}
 \label{fig:branch_compare}
\end{figure}

The addition of the dark axion portal introduces corrections to the coupling between leptons and the photon in the form of a new two-loop diagram, which changes the predicted value of the lepton anomalous magnetic moment $(g-2)_\ell$. New contributions to $a_\mu \equiv \frac{(g-2)_\mu}{2}$ are of particular interest, as the experimental measurements of the anomalous magnetic moment of the muon from Brookhaven National Laboratory \cite{Mohr:2012tt,Bennett:2002jb,Bennett:2004pv,Bennett:2006fi} exceed the best theoretical calculations by over three standard deviations \cite{PhysRevD.98.030001}: 
\begin{equation}
\Delta a_\mu = a_\mu(\text{exp}) - a_\mu(\text{SM}) = (26.8 \pm 7.6)\times10^{-10}.
\end{equation}
As was discussed in Ref. \cite{deNiverville:2018hrc}, the dark axion portal contribution is negative, and therefore aggravates the disagreement between theory and experiment.

The vector portal has the potential to explain the discrepancy between theory and experiment by introducing a new one-loop correction with a contribution equal to \cite{Pospelov:2008zw}:
\begin{equation}
    a_l^{\gamma^\prime} = \frac{\alpha}{2 \pi} \varepsilon^2 \int_0^1 dz\frac{2 m_\ell^2 z (1-z)^2}{m_\ell^2(1-z)^2+m_{\gamma^\prime}^2 z}.
\end{equation}
Values of $\varepsilon$ sufficiently large to correct the discrepancy have already been excluded by experimental searches, as shown in Fig. \ref{fig:vector_limits}.
 
\begin{figure}[t]
   \centerline{\includegraphics[width=0.48\textwidth]{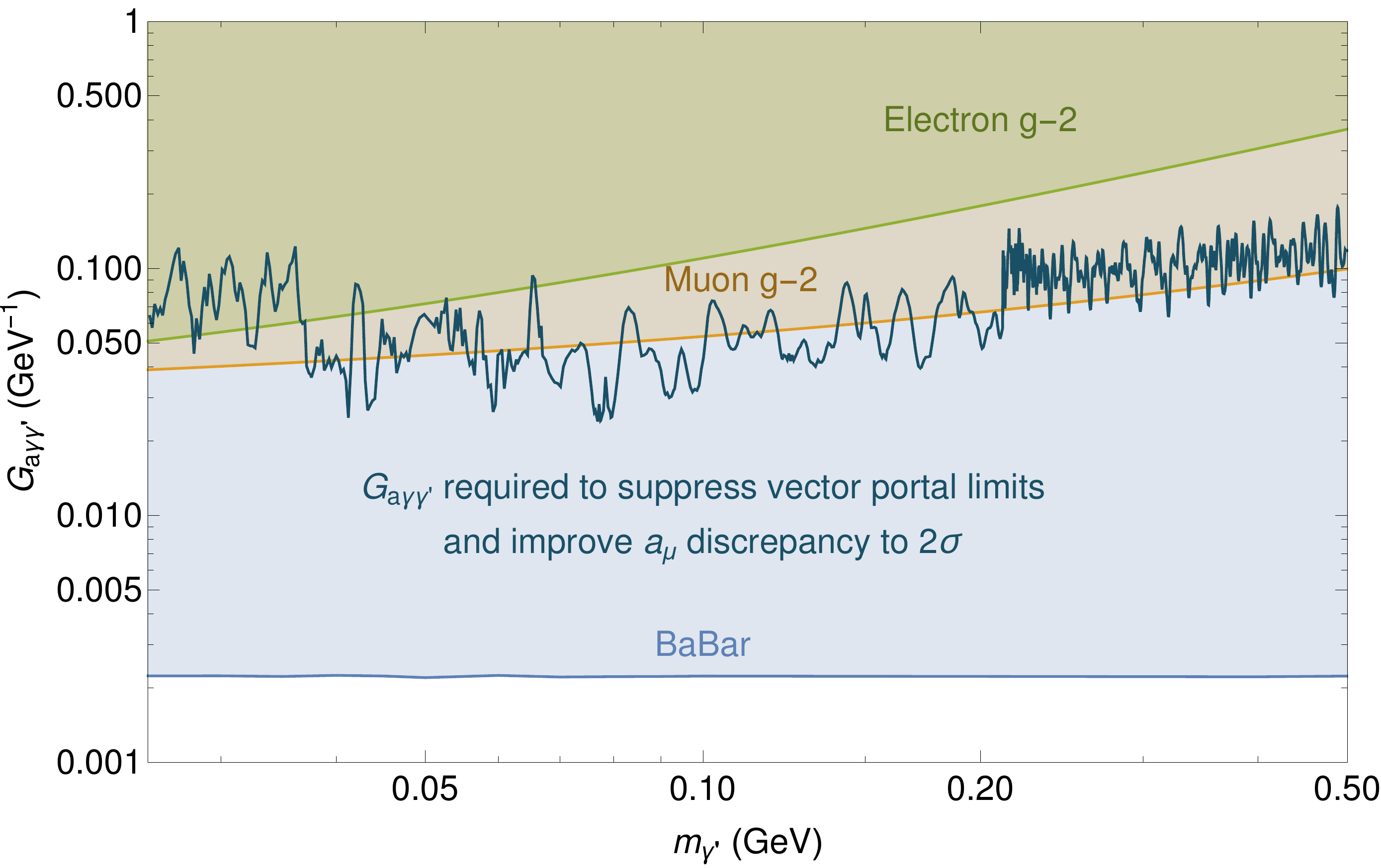}}
    \caption{The minimum value of $G_{a\gamma\gamma^\prime}$ required to suppress the vector portal limits shown in Fig. \ref{fig:vector_limits} such that the $a_\mu$ favored region is not completely excluded superimposed upon a cutout of the limits of Fig. \ref{fig:ship_limits}. The line does not extend below 25\,MeV as $a_e$ limits exclude this region. While the plot could be extended above 0.5\,GeV, its calculation is complicated by KLOE-2 searches \cite{Babusci:2012cr,Babusci:2014sta}, as they do not rely solely on the $\gamma^\prime \to e^+ e^-$ signal.}
    \label{fig:gag_limits}
\end{figure}

An intriguing possibility is that the combination of the two portals could correct the discrepancy in $(g_-2)_\mu$ while weakening the constraints placed by electron-positron decays through the vector portal. In order to accomplish this, we must satisfy three conditions:
\begin{enumerate}
 \item The kinetic mixing parameter $\varepsilon$ must be sufficiently large to make a meaningful correction to $a_\mu$, reflected in Fig. \ref{fig:vector_limits} as the blue $(g-2)_\mu$ favored band. This begins at $\varepsilon>10^{-3}$ for small $m_{\gamma^\prime}$, and grows larger with the mass.
 \item $G_{a\gamma\gamma^\prime}$ must be sufficiently small as to not be ruled out by constraints from \textsc{BABAR} or make large changes to $a_\mu$. The limits found previously in \cite{deNiverville:2018hrc} and summarized in Fig. \ref{fig:ship_limits} require that $G_{a\gamma\gamma^\prime}\lessapprox3\times 10^{-3}$.
 \item The constraints on $\varepsilon$ must be sufficiently suppressed by the introduction of new dark axion portal decay channels that the $(g-2)_\mu$ favored band in the vector portal limits plot of Fig. \ref{fig:vector_limits} is no longer completely excluded. The relevant limits for this discussion come from corrections to the electron magnetic moment $a_e$ from hidden photons \cite{Endo:2012hp}, \textsc{BABAR} \cite{Lees:2014xha}, HADES \cite{Agakishiev:2013fwl}, KLOE-2 \cite{Babusci:2012cr,Babusci:2014sta}, WASA-at-COSY \cite{Adlarson:2013eza}, A1 \cite{Merkel:2014avp}, PHENIX \cite{Adare:2014mgk} and NA48/2 \cite{Batley:2015lha}.
\end{enumerate}

Most of the vector portal limits are weakened when $G_{a\gamma\gamma^\prime}/\varepsilon$ is sufficiently large that the $\gamma^\prime$ decays preferentially through the dark axion portal process $\gamma^\prime\to a \gamma$, as this decreases Br($\gamma^\prime \to e^+ e^-$). The limit on $a_e$ is not affected by the introduction of the dark axion portal decay channel $\gamma^\prime\to a \gamma$, but may change due to corrections resulting from new dark axion portal diagrams. Finally, beam dump searches may also have their sensitivity suppressed somewhat by the reduction in the branching ratio of $\gamma^\prime\to \ell^+ \ell^-$, and may be altered further by the increase in $\Gamma_{\gamma^\prime}$ due to the introduction of new decay channels. Experiments for which the mean distance before decay is shorter than the distance between the target and decay volume will experience an exponential suppression of their sensitivity, as shown in Eq. \ref{eq:decay_event_rate}.
 
The width of $\gamma^\prime \to \ell^+ \ell^-$ through the vector portal is given by
\begin{equation}
  \Gamma(\gamma^\prime \to \ell^+ \ell^-) = \frac{\varepsilon^2 e^2}{12 \pi} m_{\gamma^\prime} \left(1-\frac{4m_\ell^2}{m^2_{\gamma^\prime}}\right)^{1/2},
  \label{eq:dec_kinetic}
\end{equation}
where $e=\sqrt{4\pi \alpha_\mathrm{em}}$. A plot of Br($\gamma^\prime \to e^+ e^-$) for different ratios of $G_{a\gamma\gamma^\prime}$ to $\varepsilon$ is shown in Fig. \ref{fig:branch_compare}. An important feature to note is that the $\gamma^\prime \to e^+ e^-$ decay channel dominates the dark photon width when $m_{\gamma^\prime}$ is small.

We can now calculate the branching ratio Br($\gamma^\prime\to e^+ e^- +X$) required for limits on the vector portal to no longer exclude the 2$\sigma$ $a_\mu$ favored region, and therefore the minimum value of $G_{a\gamma\gamma^\prime}$ required to satisfy our previous conditions. In Fig. \ref{fig:gag_limits} we plot the value of this $G_{a\gamma\gamma^\prime}$ favored line which satisfies this condition for $m_{\gamma^\prime}\in [0.025,0.5]\,$ GeV and find that these values of $G_{a\gamma\gamma^\prime}$ are easily excluded by the \textsc{BABAR} monophoton search. Worse, the coupling is sufficiently large that the dark axion portal contribution to $(g-2)_\mu$, which we recall worsens the $(g-2)_\mu$ discrepancy, is of a comparable magnitude to the vector portal contribution. For $m_{\gamma^\prime}<0.025\,\mathrm{GeV}$, limits from measurements of $a_e$ completely exclude the $(g_2)_\mu$ favored region. For $m_{\gamma^\prime}>0.5\,\mathrm{GeV}$, the vector portal limits no longer rely solely on the $\gamma^\prime \to e^+ e^- + X$ signature, and our simple treatment is no longer sufficient to calculate the $G_{a\gamma\gamma^\prime}$ favored line. The combination of the vector and dark axion portals is incapable of resolving the $(g-2)_\mu$ anomaly while satisfying existing constraints. 

\section{Dark Matter}
\label{sec:dm}

One possibility we have not considered thus far is the inclusion of an invisible dark matter candidate $\chi$ coupled to the $\gamma^\prime$. Should the coupling $g_{\gamma^\prime\chi\chi}$ be large and $m_\chi < m_{\gamma^\prime}/2$ hold, we would find that the invisible decay $\gamma^\prime \to \chi \bar \chi$ would dominate. This would heavily suppress the semivisible decays of the $\gamma^\prime$ that have comprised our strongest limits thus far. For this section, we will assume no other couplings between $\chi$ and the SM.

All of the beam dump limits considered: CHARM, SHiP and FASER, would be completely eliminated, as the $\gamma^\prime$ decays in an undetectable fashion in the unlikely event that it survives to reach one of their detectors. The \textsc{BABAR} and Belle-II monophoton limits fare rather better: While the process $e^+ e^- \to a (\gamma^\prime \to \chi \bar \chi)$ no longer provides a monophoton signal due to the rapid and invisible decay of the $\gamma^\prime$, we can instead study the rarer scattering process $e^+ e^- \to a \gamma \gamma^\prime$. These two channels were briefly compared in Fig. 3 of Ref. \cite{deNiverville:2018hrc}. In addition, the off-shell process $e^+ e^- \to a \gamma^{\prime}* \to a a \gamma$ also remains available, though highly suppressed. We would naively expect the limits from \textsc{BABAR} and Belle-II to weaken by a factor of approximately $\sqrt{20} \sim 4.5$, and without the reliance on the decay of a potentially long-lived $\gamma^\prime$, the limits would not show the same weakening behavior at low masses shown in Fig. \ref{fig:ship_limits}.

The LSND and MiniBooNE limits, on the other hand, could increase their mass reach. The fixed target neutrino experiments primarily rely on observing recoils from the scattering of the long-lived axion $a e \to \gamma^\prime e$, but their signals are complicated at high masses by the subsequent decay $\gamma^\prime \to a \gamma$ inside the detector (note that this is in and of itself an interesting new physics signature, but the currently available analyses cannot account for it). With the inclusion of the invisible decay of the $\gamma^\prime$, the limits placed by these experiments can extend to tens of MeV. The strength of the limit in $G_{a\gamma\gamma^\prime}$ will be weakened slightly, as the $\gamma^\prime$ particles produced in the target no longer serve as a secondary source of $a$ particles through their subsequent decays.

The limits from lepton $g-2$ are largely unaffected by the introduction of dark matter particles $\chi$. We can, however, reexamine the arguments of the previous section by considering a combination of the dark axion portal, the vector portal and the coupling $G_{\gamma^\prime\chi\chi}$, and come to the same conclusions.The dark axion portal cannot resolve the $g_\mu-2$ discrepancy on its own, and the inclusion of dark matter does not change matters, forcing us to consider the effect on the vector portal. While it is true that this new decay channel heavily suppresses or completely eliminates all of the beam dump and rare decay limits shown in Fig. \ref{fig:vector_limits}, NA64 \cite{NA64:2019imj} and \textsc{BABAR} \cite{Lees:2017lec} analyses place strong constraints on a kinetically mixed dark photon that decays invisibly. These limits completely rule out the kinetic mixing regime where $g_\mu-2$ might be improved by the inclusion of a dark photon. 

\section{Monophoton Search}
\label{sec:monophoton}

\begin{figure}[t!]
 \centerline{\includegraphics[width=0.5\textwidth]{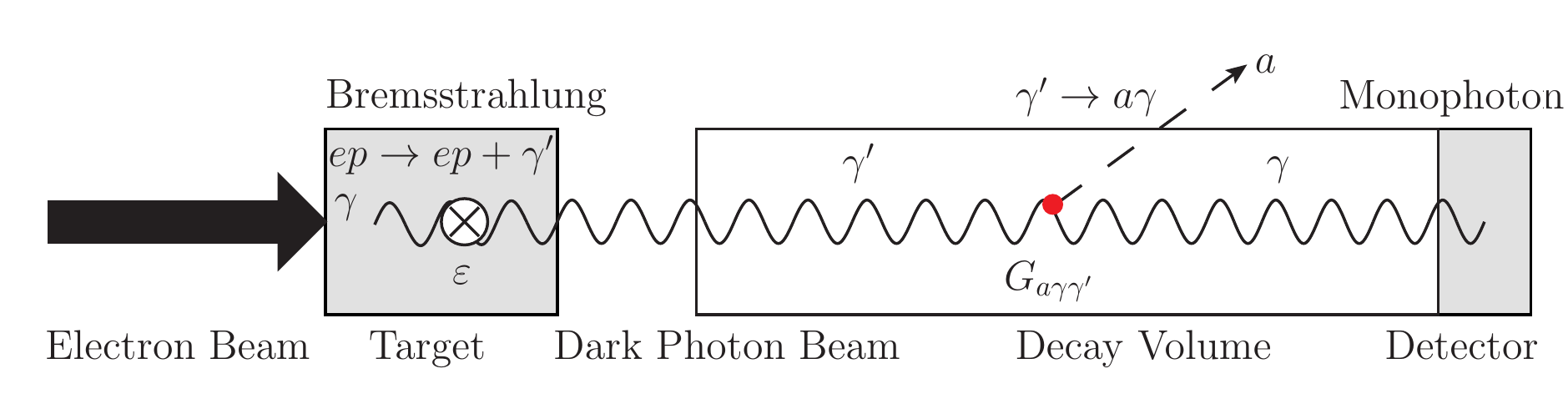}}
 \caption{A schematic of an electron beam dump experiment searching for monophoton signals. A scenario of particular interest is one for which dark photon production proceeds through vector portal bremsstrahlung, while the decay to $a\gamma$ proceeds through the dark axion portal. The photon is observed in the detector as a monophoton, while the invisible $a$ escapes. A possible variant involves searching for the decay of a sufficiently short-lived $a$, resulting in a three-photon signal produced through the axion portal decay $a\to \gamma\gamma$.}
 \label{fig:elec_beam}
\end{figure}

\begin{figure}[t]
\subfigure[]{ 
\centerline{\includegraphics[width=0.45\textwidth]{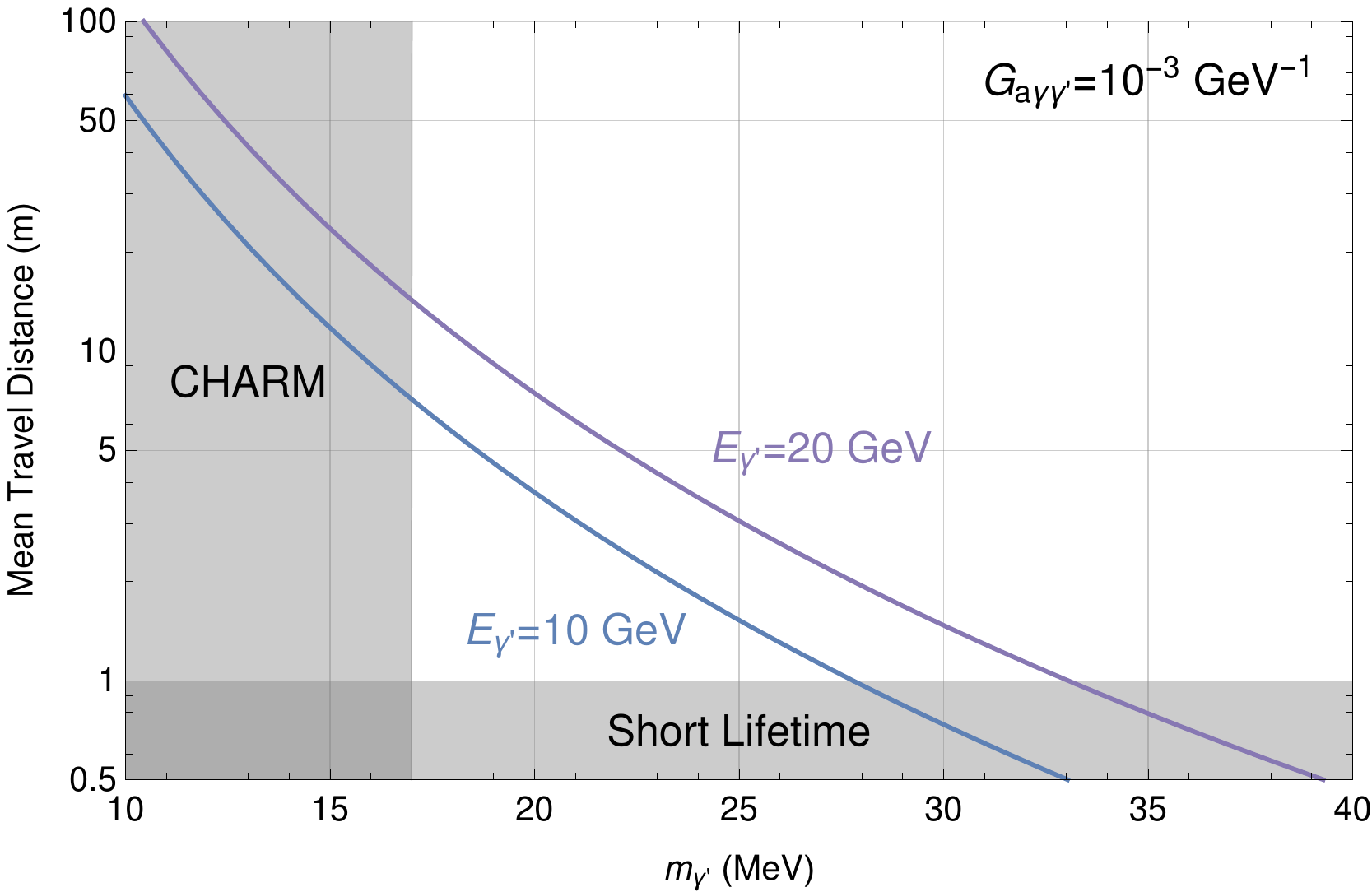}}
}
\subfigure[]{
\centerline{\includegraphics[width=0.45\textwidth]{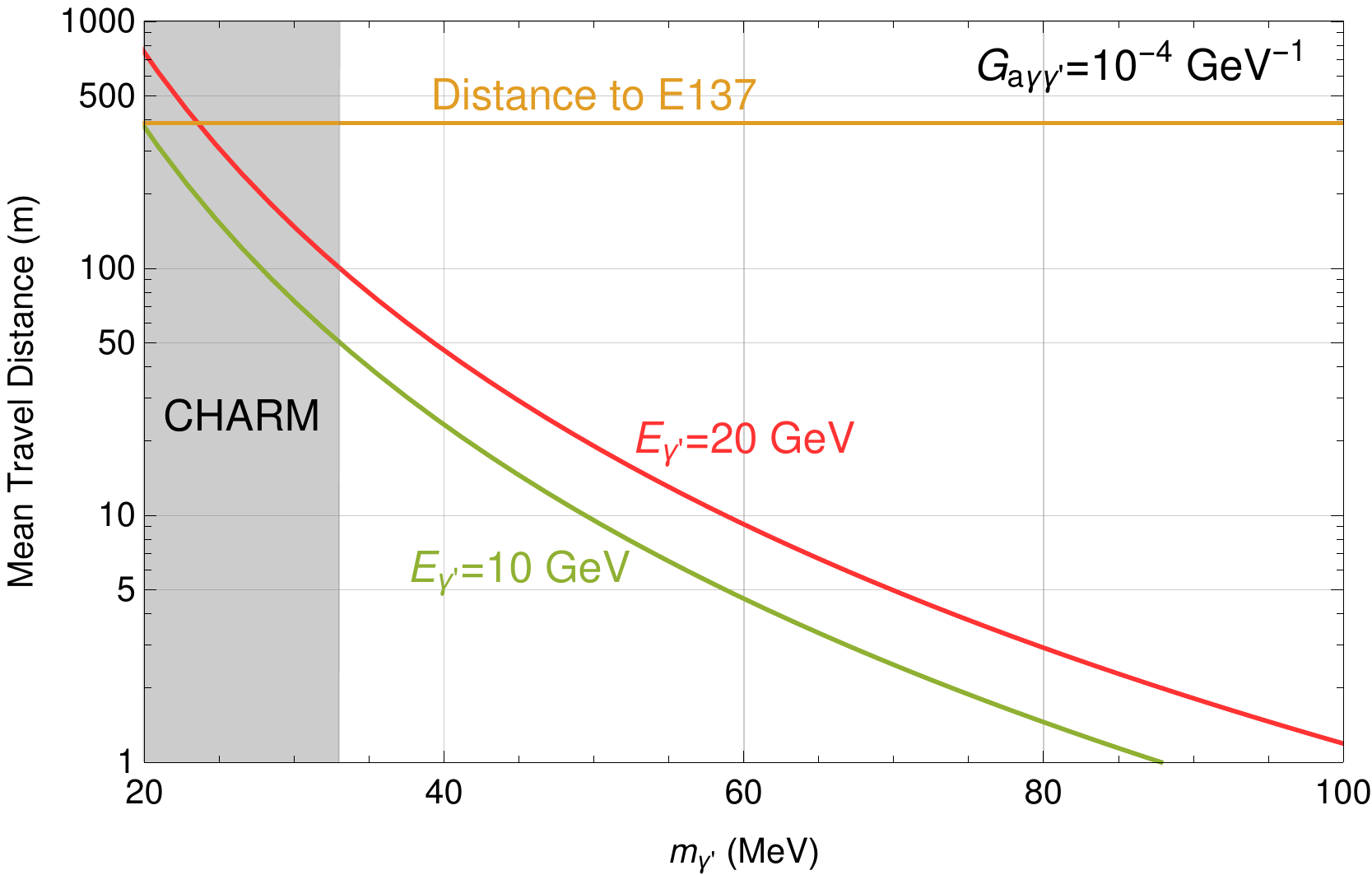}}
}
\caption{The mean travel distance before decay of a $\gamma^\prime$ with energy equal to 10 and 20 GeV for two interesting values of $G_{a\gamma\gamma^\prime}$. Dark photons produced through electron bremsstrahlung are highly collimated with the beam direction and possess an energy nearly equal to that of the electron beam itself. Parameters for which the decay length is much shorter than a meter are likely to decay before escaping the target and any shielding present, while extremely large decay lengths many times larger than the detector size result in very few of the dark photon decays inside of the detector. The target-detector distance for E137, a comparable electron beam dump experiment with a beam energy of 20\,GeV, is shown in (b) for reference.}
\label{fig:travel_dist}
\end{figure}

The previous section considered the effects of combining the vector and dark axion portals but ignored the possibility that there may be strong new experimental constraints on a scenario combining both portals. Frequently, one portal will dominate over the other in interactions with the SM, but this need not always be the case, and for some combinations, production may occur through one portal, while detection proceeds through another. 

The production rate of $\gamma^\prime$ and $a$ is suppressed in the dark axion portal by the need to produce the particles in pairs, resulting in three-body final states or worse. If vector portal processes instead produced the dark photons, the production rate could be greatly enhanced at the cost of satisfying the robust existing constraints on the kinetic mixing parameter $\varepsilon$. We can suppress the limits on a kinetically mixed dark photon if the dark axion portal decay $\gamma^\prime \to a \gamma$ dominates over leptonic decay modes. As was shown in Sec. \ref{sec:muon_g_2}, this requires $G_{a\gamma\gamma^\prime}\gg \varepsilon/\mathrm{GeV}$, which is possible due to the comparatively weak constraints on the dark axion portal, and that $m_{\gamma^\prime}$ be greater than few tens of MeV. The latter requirement is due to the strong $m_{\gamma^\prime}$ dependence of the branching ratio shown in Fig. \ref{fig:branch_compare}. There is still the potential for an improvement over the proton beam dump limits we have considered previously on a timescale shorter than that required by SHiP, particularly from electron beam dump experiments with dark photon bremsstrahlung as a production channel as suggested in Fig. \ref{fig:elec_beam}.

Electron beam dumps could supplement the exclusion regions already provided by \textsc{BABAR} and CHARM, though their exact reach is dependent on the magnitude of the kinetic mixing parameter $\varepsilon$. A beam dump experiment's sensitivity is at its greatest when the mean decay length is equal to the distance between the decay volume and the production target. This distance is dependent on both the coupling through which the decay proceeds and the energy of the decaying particle. For electron bremsstrahlung, we can use the improved Weizsäcker-Williams approximation \cite{Kim:1973he,Tsai:1973py,Tsai:1986tx} to find that the emitted dark photon will possess an energy nearly equal to that of the electron beam itself so long as the electron beam energy is far greater than the electron mass \cite{Bjorken:2009mm,Liu:2017htz,Gninenko:2017yus}.  We consider the mean travel distance for both 10 and 20 GeV electron beams for $G_{a\gamma\gamma^\prime}=10^{-3},10^{-4}$\,GeV$^{-1}$ in Fig. \ref{fig:travel_dist} to illustrate the regions of greatest potential reach for such an experiment. Of particular interest is the $m_{\gamma^\prime}<100\,\mathrm{MeV}$ region below the existing CHARM limit. As an example, the E137 experiment may be capable of placing new limits, but with a target-to-detector distance of 383\,meters, it is likely only to be sensitive to the $m_{\gamma^\prime}<30\,\mathrm{MeV}$ region already covered by CHARM. An electron beam dump could make a significant improvement on these experiments with a shorter target-detector distance on the order of tens of meters and the ability to record monophoton events. Note that the limits on the dark photon from vector portal couplings are particularly strong for $m_{\gamma^\prime}$ of a few MeV, as the electron-positron decay channel dominates for these masses, as shown in Fig. \ref{fig:branch_compare}.

\section{Summary and Discussion}
\label{sec:summary}

We calculated the sensitivity of the proposed SHiP, FASER and FASER 2 experiments to the dark axion portal in the $m_a \ll m_\gamma^\prime$ limit, and summarized them in Fig. \ref{fig:ship_limits}. We examined the effects of combining the vector portal and dark axion portal on the anomalous magnetic moment of the muon, and building on this we suggested that the two portals could be combined to search for kinetically mixed dark photons with monophoton decay signatures produced through $\gamma^\prime \to a \gamma$. 

SHiP is capable of great improvements over the existing beam dump limits from CHARM by searching for the dark photon decay signature $\gamma^\prime\to e^+ e^- a$ using the same selection cuts as the planned vector portal dark photon search under the assumption of zero backgrounds. The large boost and short baseline provide SHiP with sensitivity to values of $m_{\gamma^\prime}$ as large as 250\,MeV. A monophoton search could further improve SHiP's sensitivity, with particularly large gains for $m_{\gamma^\prime} \sim 2 m_e$, though the exact enhancement is heavily dependent on the severity of $\gamma$ related background. Figure \ref{fig:ship_mono} shows a possible 100-event contour. 

FASER and FASER 2 were also studied, and provide an interesting contrast with SHiP. The much lower collision rate between protons and smaller length of decay volume relative to the interaction point-detector distance severely weakens FASER's sensitivity to the dark axion portal. FASER 2's possesses greatly improved sensitivity due to its much larger size and available luminosity significantly improves its sensitivity, allowing it to probe regions of the parameter space unavailable to CHARM. We briefly considered the benefits of a monophoton search at FASER 2, but it may be challenging to overcome SM backgrounds.

We calculated the effect on $(g-2)_\mu$ of the combined dark axion and vector portals in an attempt to resolve the discrepancy between theory and experiment. We found that by introducing the $\gamma^\prime \to a \gamma$ decay channel the constraints on the kinetic mixing $\varepsilon$ could be suppressed, freeing the $(g-2)_\mu$ favored region of the parameter space shown in Fig. \ref{fig:vector_limits} from existing limits. However, the values of $G_{a\gamma\gamma^\prime}$ required to sufficiently suppress the constraints on $\varepsilon$ were excluded by limits from \textsc{BABAR}.

Novel searches are possible for the combined vector and dark axion portal scenario. One possibility is the production of dark photons through bremsstrahlung and decay to a monophoton through $\gamma^\prime \to a \gamma$. The combination of portals can take advantage of the numerous electron beam dump experiments that would otherwise lack sensitivity to the dark axion portal, but the absence of monophoton analyses may hamper their potential searches. A short-baseline electron beam dump experiment with sensitivity to monophoton signals would provide an ideal environment for this kind of search.

\begin{acknowledgments}
This work was supported in part by IBS (Project Code IBS-R018-D1) and NRF Strategic Research Program (NRF-2017R1E1A1A01072736). We would like to thank Walter Bonivento for helpful discussions regarding the SHiP experiment. We also would like to thank the participants of the Light Dark World International Forum 2018 (KAIST, December 2018) for engaging discussions.
\end{acknowledgments}

\bibliography{DAP_ShipFaser}

\end{document}